\documentclass[10pt, conference, compsocconf]{IEEEtran}
\usepackage{cite}
\usepackage{url}
\usepackage{amssymb}

\newcommand{\poista}[1]{}

\newtheorem{theorem}{Theorem}

\newtheorem{lemma}[theorem]{Lemma}

\newcommand{\tb}{\textbf}

\newcommand{\ts}{\textsf}
\newcommand{\ttt}{\texttt}
\newcommand{\mc}{\mathcal}
\newcommand{\mi}{\mathit}
\newcommand{\ms}{\mathsf}

\newcommand{\ar}[1]{\,{-}#1{\rightarrow}}
\newcommand{\arr}[1]{\,{-}#1{\rightarrow}\,}

\usepackage{pstricks}
\psset{xunit=1pt,yunit=1pt,runit=1pt,linewidth=0.5pt}
\newcommand{\psst}{\pscircle[fillstyle=solid,fillcolor=white]}

\newcommand{\arc}{\psline[linearc=2]{->}}

\newcommand{\ainit}[1]{\rput[lb](#1){\arc(-4,8)(-0.9,1.8)}}

\begin{document}

\title{Stop It, and Be Stubborn!}

\author{\IEEEauthorblockN{Antti Valmari}
\IEEEauthorblockA{Department of Mathematics\\
Tampere University of Technology\\
Tampere, Finland\\
Email: Antti.Valmari@tut.fi}}

\maketitle

\begin{abstract}
A system is \tb{AG} \tb{EF} terminating, if and only if from every reachable
state, a terminal state is reachable.
This publication argues that it is beneficial for both catching non-progress
errors and stubborn set state space reduction to try to make verification
models \tb{AG} \tb{EF} terminating.
An incorrect mutual exclusion algorithm is used as an example.
The error does not manifest itself, unless the first action of the customers
is modelled differently from other actions.
An appropriate method is to add an alternative first action that models the
customer stopping for good.
This method typically makes the model \tb{AG} \tb{EF} terminating.
If the model is \tb{AG} \tb{EF} terminating, then the basic strong stubborn
set method preserves safety and some progress properties without any
additional condition for solving the ignoring problem.
Furthermore, whether the model is \tb{AG} \tb{EF} terminating can be checked
efficiently from the reduced state space.
\end{abstract}


\section{Introduction}\label{S:intro}

Reduced state space construction using stubborn set / partial order methods is
based on, in each constructed state, computing a subset of transitions and
only firing the enabled transitions in it instead of all enabled transitions.
This set is known under various names, such as stubborn
set~\cite{Val88,Val98}.

The problem of how to compute stubborn sets is non-trivial.
Often it is solved in two steps.
First, a collection of abstract conditions is presented such that if the sets
conform to them, then the reduced and full state spaces yield the same values
of certain correctness properties.
A widely known example of such a collection are D0, D1, and D2 in
Section~\ref{S:background}.
The choice of the conditions depends on the chosen correctness properties.
In an attempt to improve reduction results, various conditions have been
suggested even for the same properties.
 
Second, concrete algorithms are presented that construct sets that obey the
conditions.
They are not discussed further here, because the present publication does not
aim at contributing to them.

The conditions D0, D1, and D2 constitute the \emph{basic strong stubborn set
method}.
It guarantees that the full and reduced state spaces have precisely the same
terminal states and that the reduced state space has an infinite execution if
and only if also the full state space has.
Because of the so-called \emph{ignoring problem}, it does not guarantee to
preserve safety properties such as mutual exclusion and liveness properties
such as eventual access.
That is, the method may investigate a part of the state space that is
unimportant for the property and then stop.

To preserve safety properties, various additional conditions have been
suggested.
One possibility is to recognize the terminal strong components of the reduced
state space and ensure that in each of them, every enabled transition is
fired.
To preserve liveness properties, a common strategy is to ensure that every
enabled transition is fired in every cycle of the reduced state space.

These additional conditions are problematic in two respects.
First, there is the general phenomenon that the more conditions there are, the
more enabled transitions the stubborn sets contain, and the bigger the reduced
state space becomes.
Second, as was pointed out in~\cite{EvP10}, a condition may choose the states
where it fires all enabled transitions in an unfortunate way, leading to the
construction of many more states than would be needed.
The well-known liveness condition in~\cite{CGP99} suffers from this problem.

In the present publication, a stunningly simple solution to the ignoring
problem is suggested, proven correct, and experimented with.
It suffices for safety and some progress properties.
It is: if the modeller tries to make the verification models \emph{\tb{AG}
\tb{EF} terminating}, then no additional conditions are needed at all.
A model is \tb{AG} \tb{EF} terminating if and only if from every reachable
state, a terminal state is reachable.
The notation ``\tb{AG} \tb{EF}'' comes from the well-known logic called
CTL~\cite{Eme90}.
``Tries to make'' refers to the fact that the modeller need not prove that the
model is \tb{AG} \tb{EF} terminating.
Instead, the model checker tool checks whether it is.
In other words, not being \tb{AG} \tb{EF} terminating is considered an error,
and the model checker is guaranteed to reveal it (unless another error stops
it first).

Trying to make models \tb{AG} \tb{EF} terminating is a more natural goal than
it might first seem.
Using Peterson's mutual exclusion algorithm for $n$ customers~\cite{Pet81} as
an example,
Section~\ref{S:example} demonstrates that naive modelling may lead to the loss
of non-progress errors.
It is justified in Section~\ref{S:appropriate} that this problem can be solved
by making the customers of the algorithm capable to choose to terminate.
This makes the model as a whole \tb{AG} \tb{EF} terminating.
That is, even forgetting about stubborn sets, to check the eventual access
property, the model must be made \tb{AG} \tb{EF} terminating (or some more
complicated method such as a suitable weak fairness assumption must be used).

A counterexample in~\cite{Val98} leaves little hope of finding an essentially
better condition for the full class of linear-time liveness properties than
some variant of the cycle condition.
Furthermore, the validity of linear-time liveness properties often depends on
so-called fairness assumptions.
They may be problematic for the modeller (have you ever tried to teach them to
software engineering students?), and they have never been integrated well to
stubborn sets.
A series of counterexamples illustrating the difficulties was prepared for
this publication, but was left out because of lack of space.

In essence, linear-time liveness corresponds to the formula \tb{AF} $\varphi$,
where $\varphi$ denotes something desirable such as receiving service, and the
formula is stipulated on the states where the service is requested.
Some authors have advocated the use of the strictly weaker notion \tb{EF}
$\varphi$.
For instance, a process-algebraic variant of this theme was presented
in~\cite{ReV07}.
With it, fairness assumptions become unnecessary.
It may or may not be a sufficiently stringent correctness property from the
practical point of view, but certainly it is much better than nothing.

With \tb{AG} \tb{EF} terminating systems, this weaker notion reduces to the
requirement that no terminal state has unsatisfied service requests.
This condition can be modelled as a check that is run on each terminal state.
The present author believes that modellers will not find it difficult to
formulate such checks.

This approach facilitates early on-the-fly detection of safety and
non-progress errors.
The basic stubborn set method suffices during state space reduction.
The order in which the state space is constructed is left unspecified, making
it possible to use breadth-first for short counterexamples.
As described in Section~\ref{S:new-stubborn}, the check that the system indeed
is \tb{AG} \tb{EF} terminating can be implemented as a postprocessing step
that is performed only if no errors are revealed during state space
construction.

In addition to the \tb{EF} $\varphi$ approach to progress, a subset of
linear-time liveness properties is covered in Section~\ref{S:new-stubborn}.

Section~\ref{S:background} presents the necessary background on stubborn sets.
The new theorems are developed in Section~\ref{S:new-stubborn}.
Experimental results obtained with a new state space tool that implements the
approach are reported in Section~\ref{S:conclusions}.
Section~\ref{S:stubbrule} describes the choice of the stubborn sets in the
experiments.

\section{A Motivating Example}\label{S:example}

Consider one or more concurrent processes called \emph{customers}, each of
which has a distinguished piece of code called \emph{critical section}.
The purpose of a \emph{mutual exclusion algorithm} is to ensure that at any
instant of time, no more than one customer is in the critical section.
The algorithm must have the \emph{eventual access property}, that is, if any
customer tries to enter the critical section, it eventually succeeds.
Typically it is assumed that an atomic operation can access at most one shared
variable, and only once.
For instance, if \ttt{i} is a shared variable, then \ttt{++i} involves at
least two atomic operations, one reading the original value of \ttt{i} and
another writing the new value.

\begin{figure}
/* protocols for $P_i$ */\\
\tb{for} $j$ := $1$ \tb{to} $n-1$ \tb{do}\\
\tb{begin}\\
\mbox{} ~ $Q[i]$ := $j$;\\
\mbox{} ~ $\mi{TURN}[j]$ := $i$;\\
\mbox{} ~ \tb{wait until} ($\forall k \neq i, Q[k] < j$) \tb{or} $\mi{TURN}[j]
\neq i$\\
\tb{end};\\
Critical Section;\\
$Q[i]$ := $0$
\caption{Peterson's algorithm for $n$ customers~\cite{Pet81}.}\label{F:Pete-n}
\end{figure}
\emph{Peterson's algorithm} is a famous algorithm for solving the mutual
exclusion problem on this level of atomicity.
In his original publication~\cite{Pet81}, Peterson first described his
algorithm for two customers and then generalized it to $n$ customers for an
arbitrary fixed positive integer $n$.
We call these algorithms ``Peterson-two'' and ``Peterson-$n$'', respectively.

Figure~\ref{F:Pete-n} shows Peterson-$n$ copied verbatim from~\cite{Pet81}.
It implements $n-1$ \emph{gates}.
To go through gate $j$, customer $i$ writes $j$ to the shared variable $Q[i]$.
Then it gives priority to other customers by writing its own number $i$ to the
$\mi{TURN}$ variable of the gate.
It can go through the gate when no other customer is trying to go through the
same or further gate, or when some other customer comes to the same gate,
changing the $\mi{TURN}$.

\begin{figure*}{\small
\begin{verbatim}
const unsigned n = 3;   // number of customers

state_var
  S[n],   // state of customer i: 0 = idle, 7 = critical, 1...6 = trying
  j[n],   // local variable j of customer i
  k[n],   // local variable k of customer i
  Q[n],   // number of gate through which customer i wants to go
  T[n-1]; // number of customer who has _no_ priority at gate j

const char ltr[] = { '-', 'j', 'Q', 'T', 'w', 'k', 'A', '*' };
void print_state(){
  for( unsigned i = 0; i < n; ++i ){
    std::cout << j[i] << ltr[ S[i] ] << k[i] << Q[i] << ' ';
  }
  for( unsigned i = 0; i < n-1; ++i ){ std::cout << T[i]; }
  std::cout << '\n';
}

/* Check that at most one customer is in critical section at any time */
#define chk_state
const char *check_state(){
  unsigned cnt = 0;
  for( unsigned i = 0; i < n; ++i ){ if( S[i] == 7 ){ ++cnt; } }
  if( cnt >= 2 ){ return "Mutex violated"; }
  return 0;
}

/* Check that customer 0 may always make progress. */
#define chk_may_progress
bool is_may_progress(){ return S[0] == 7; }

unsigned nr_transitions(){ return n; }

bool fire_transition( unsigned i ){
  #define goto(x){ S[i] = x; return true; }
  switch( S[i] ){
  case 0: j[i] = 0; goto(1)
  case 1: if( j[i] >= n-1 ){ goto(7) }else{ goto(2) }
  case 2: Q[i] = j[i]; goto(3)
  case 3: T[j[i]] = i; goto(4)
  case 4: if( T[j[i]] != i ){ ++j[i]; goto(1) }else{ k[i] = 0; goto(5) }
  case 5: if( k[i] >= n    ){ ++j[i]; goto(1) }else{ goto(6) }
  case 6: if( k[i] == i || Q[k[i]] < j[i] ){ ++k[i]; goto(5) }else{ goto(4) }
  case 7: /* critical section */  Q[i] = 0; goto(0)
  default: err_msg = "Illegal local state"; return false;
} }
\end{verbatim}}
\caption{A questionnable model of Peterson's algorithm for $n$
customers.}\label{F:Pete-q}
\end{figure*}
Figure~\ref{F:Pete-q} shows a model of Peterson-$n$ written for ASSET.
ASSET is A State Space Exploration Tool that is based on presenting the model
as a collection of C++ functions that obey certain conventions~\cite{Val15}.
The model is checked by copying it to the file \ttt{asset.model} and then
compiling and executing \ttt{asset.cc}.
This approach facilitates very fast execution of the transitions of the model
and makes the modelling very flexible, because most features of C++ are
available.
On the other hand, the modelling language does not always support intuition
well.
This problem could be solved by implementing a preprocessor tool that inputs
some nice modelling language and outputs the input language of ASSET.
At the time of writing, no such tool has been implemented.

Variables that describe the state of the model must be of the special type
\ttt{state\_var}.
The value of such a variable is an unsigned integer in the range $0, \ldots,
2^b-1$, where $b=8$ by default but can be specified for each state variable or
array of state variables individually.

The modelling of the shared array $Q$ of Peterson-$n$ is obvious in
Figure~\ref{F:Pete-q}.
The shared array $\mi{TURN}$ has been abbreviated to \ttt{T}.
Because the input language of ASSET has no notion of local variables of
processes, $j$ and $k$ have been modelled as arrays.
That is, \texttt{j[$i$]} models the $j$ of customer $i$, and similarly with
\texttt{k[$i$]}.
The variable \ttt{S[$i$]} keeps track of the local state of customer $i$.
It can be thought of as a program counter.

When ASSET has found an error, it prints a counterexample in the form of a
sequence of states that leads from the initial state to an error state.
Depending on the type of the error, the counterexample may also contain a
cycle of states where the system fails to make progress towards some desired
situation.
For this purpose, the model must contain a \ttt{print\_state} function.
The function in Figure~\ref{F:Pete-q} presents the local states of the
customers as characters, to make the print-out easier to interpret.
Other than that, the function is straightforward.

The next function specifies the mutual exclusion property.
The line \ttt{\#define} \ttt{chk\_state} commands ASSET to check every state
that it has found by calling the \ttt{check\_state} function.
(It would have been nicer to use the same word \ttt{check\_state} both in
\ttt{\#define} and as the function name.
Unfortunately, C++ does not allow that.)
By returning a character string, the function indicates that something is
wrong with the state.
This makes ASSET terminate the construction of the state space and print an
error message that contains the string.
That the state is good is indicated by returning the null pointer \ttt{0}.

The line \ttt{\#define} \ttt{chk\_may\_progress} and the function immediately
after it specify that for every state that the model can reach, the model may
continue to a state where customer 0 is in the critical section.
That is, the model cannot go into a state from which there is no path to a
state where customer 0 is in the critical section.
This represents the eventual access property.
We call this particular form ``may-access''.

Peterson-two satisfies a stronger eventual access property which we call
``must-access''.
In it, after a customer has set its $Q$ variable, every path in the state
space eventually leads to a state where the customer is in the critical
section.
However, because the ``$\forall k \neq i, Q[k] < j$'' test in Peterson-$n$
accesses more than one shared variable, and because one atomic operation may
access at most one shared variable, the test must be implemented as a loop.
An unsuccessful test introduces a cycle in the state space that does not take
the customer to the critical section.
That is, Peterson-$n$ does not satisfy must-access.
If must-access is specified and there are at least two customers, then ASSET
reports an error.
This is why may-access is used in Figure~\ref{F:Pete-q}.

ASSET calls the function \ttt{nr\_transitions} to find out how many
transitions the model contains.
The model in Figure~\ref{F:Pete-q} has one transition for each customer.
It models all atomic operations of the customer.
The grouping of atomic operations to transitions for ASSET is rather flexible.
The only strict rule is that if two atomic operations may be executed in the
same state and they yield different states, then they must belong to different
transitions.
This is because for ASSET, transitions must be deterministic.
(This implies that a nondeterministic atomic operation must be modelled with
more than one transition.)

Finally, the function \ttt{fire\_transition} specifies the transitions.
Given the number of a transition, it must either return \ttt{false} indicating
that the transition is disabled in the current state, or modify the state
according to the effect of the execution of the transition and return
\ttt{true}.
If it returns \ttt{false}, then it must not modify the state.

To improve readability, Figure~\ref{F:Pete-q} introduces a \ttt{goto(x)}
macro.
It moves the customer to local state \ttt{x} and indicates that the transition
was enabled.

The modelling of the atomic operations $Q[i]$ := $j$, $\mi{TURN}[j]$ := $i$,
and $Q[i]$ := $0$ of Peterson-$n$ is straightforward.
The \tb{for} $j$ loop has been modelled by cases 0 and~1 and the \ttt{++j[i];}
\ttt{goto(1)} in cases 4 and 5.
In Figure~\ref{F:Pete-n}, arrays are indexed starting from 1.
However, consistently with the usual C++ convention, indexing starts from 0 in
Figure~\ref{F:Pete-q}.
Cases~4 to 6 model the line
\begin{center}
\tb{wait until} ($\forall k \neq i, Q[k] < j$) \tb{or} $\mi{TURN}[j] \neq i$
\end{center}
with two loops.
The inner loop tries $k = 0$, $k = 1$, and so on until the $\forall$ test is
found to fail or pass.
The outer loop first tests whether $\mi{TURN}[j] \neq i$ and if it fails, then
starts the inner loop.
If either test passes, then \ttt{++j[i];} \ttt{goto(1)} is executed,
modelling the completion of an iteration of the \tb{for}~$j$ loop.
If the $\forall$ test fails, then the outer loop is started again.
Because the model checking is obviously incomplete if each customer tries only
once to go to the critical section, a jump to the beginning has been added to
case 7.
The \ttt{default} branch is never entered, but the compiler complains if it is
absent.

The entry ``plain not non-progress revealing'' in Table~\ref{T:result} in
Section~\ref{S:conclusions} shows the state space size and state space
construction and exploration time in seconds of this model.
ASSET reported no errors.
Excluding small variation in the times, the results remained the same when any
customer replaced customer 0 in \ttt{is\_may\_progress}.
However, the model was deemed ``questionnable'' in the caption of
Figure~\ref{F:Pete-q}.
The reason for this will be discussed next.

In a second model of Peterson-$n$, customers were made able to not try to go
to the critical section.
To do this, a new local state 8 and a new transition to it from local state 0
were added to the model.
When in local state~0, the customer chooses nondeterministically and without
being affected by the other customers to either terminate (by going to local
state 8) or to start trying to go to the critical section (by executing
\ttt{j[i]} \ttt{=} \ttt{0;} and going to local state 1).
This was implemented by adding \ttt{'~'} to the end of \ttt{ltr}; making
\ttt{is\_may\_progress} to return \ttt{true} if and only if \ttt{S[0]}
\ttt{>=} \ttt{7}; making \ttt{nr\_transitions} return \ttt{2*n}; adding
\ttt{case} \ttt{8:} \ttt{return} \ttt{false;} to the \ttt{switch} statement;
and adding the following lines immediately before the \ttt{switch} statement:
\begin{verbatim}
if( i >= n ){
  i -= n;   // extract customer number
  if( S[i] == 0 ){ goto(8) }
  return false;
}
\end{verbatim}

When $n=2$, ASSET gives the following error message after 2.3\,s of
compilation and negligible analysis time:
\begin{verbatim}
0-00 0-00 0
0-00 0 00 0
==========
0j00 0 00 0
0Q00 0 00 0
0T00 0 00 0
0w00 0 00 0
----------
0k00 0 00 0
0A00 0 00 0
0k10 0 00 0
0A10 0 00 0
0w10 0 00 0
!!! May-type non-progress error
163 states, 326 arcs
\end{verbatim}
In it, customer 1 terminates (its local state changes from ``\ttt{-}'' to
``\ttt{~}'') and then customer 0 goes to local state 1 (``\ttt{j}'').
The line \ttt{==========} indicates that there is no path from this state to a
state where customer~0 is in local state 7 or 8.
This means that \emph{customer~0 cannot go to the critical section after
customer 1 has terminated}.
So eventual access fails even in its less stringent form ``may-access''.

In the continuation of the counterexample, customer 0 goes to local state 4 as
expected.
Then it starts to run around a loop where it first executes cases 5 and 6 with
$\ttt{k} = 0$, then it executes them with $\ttt{k} = 1$, and then goes back to
local state 4.
This corresponds to being stuck in the statement
\begin{center}
\tb{wait until} ($\forall k \neq i, Q[k] < j$) \tb{or} $\mi{TURN}[j] \neq i$
\end{center}

Why does the customer not go through the gate?
The part $\mi{TURN}[j] \neq i$ does not let it pass, because it can be seen
from the state that $\ttt{T[}0\ttt{]} = 0$.
Neither does the $\forall$ part, because $\ttt{i} = 0$, $\ttt{Q[}1\ttt{]} =
0$, and $\ttt{j[}0\ttt{]} = 0$.
In Peterson-$n$, $Q[1]$ would be $0$ but $j$ would be $1$, because the
\tb{for}-loop starts with $j=1$ in it.
In Figure~\ref{F:Pete-q} indexing and thus also the \tb{for}-loop were made to
start with $j=0$.
This made the value $0$ in entries of $Q$ incorrectly mean both that the
customer is not trying to go to the critical section and that it is trying
to go through or has just gone through the first gate.
It looks to customer 0 like the terminated customer 1 were trying to go
through the first gate.
Because customer 0 does not have priority, it keeps on running around the
waiting loop.

Why does the error not manifest itself in the original model?
In it, if customer $i_1$ is waiting at the first gate, some other customer
$i_2$ eventually arrives at the same gate.
This is because always at least one customer can make progress (the \ttt{T}
test blocks at most one customer per gate and there are one fewer gates than
customers), a progressing customer eventually reaches local state 0, and when
all progressing customers are there, there is nothing else that the model can
do than move one of them to the first gate.
When arriving there, it gives priority to customer $i_1$ by assigning $i_2$ to
\ttt{T[0]}.

To fix the error, \ttt{Q[i]} \ttt{=} \ttt{j[i]} in case 2 was changed to
\ttt{Q[i]} \ttt{=} \ttt{j[i]+1} and the latter test in case 6 was changed to
\ttt{Q[k[i]]} \ttt{<=} \ttt{j[i]}.
The entry ``correct'' in Table~\ref{T:result} shows the results with the fixed
model.
No matter which customer was tested by \ttt{is\_may\_progress}, ASSET reported
no errors.

To have an example of safety errors, finally the model was analysed that was
obtained by swapping the statements \ttt{Q[i]} \ttt{=} \ttt{j[i]+1;} and
\ttt{T[j[i]]} \ttt{=} \ttt{i;} in the correct model.
ASSET reported a mutual exclusion error.
In it, customer 1 went through the first gate while customer 0 stayed between
the above-mentioned statements.
Customer~1 got through, because \ttt{Q[0]} had not yet been assigned to.
Then customer~0 got through because when customer 1 had passed by, it had
given priority to customer 0 by assigning \ttt{1} to \ttt{T[0]}.

\section{Appropriate Modelling of Not Requesting}\label{S:appropriate}

The message of Section~\ref{S:example} is that
\begin{quote}
\emph{If, in a model, customers are not made able to not request for service,
then non-progress errors may escape model checking.}
\end{quote}

Another way to look at this is that the first line in Figure~\ref{F:Pete-n} is
different from the other places in that while in any of the latter, the
customer must eventually go further if it can.
Why the same must not be required of the first line was found out in
Section~\ref{S:example}.
It is obvious that if a customer never leaves the critical section, then the
eventual access property cannot be provided to other customers without
violating mutual exclusion.
Although it is less obvious, a similar argument applies to most other places.
For the remaining places the requirement is at least reasonable, even if not
absolutely necessary.

In linear temporal logic~\cite{MaP92}, the customary way to express this
difference is to assume so-called weak fairness towards all other transitions
but not towards those that model the customers making requests.
Every model has an imaginary ``idling'' transition that is always enabled.
When the only thing that the modelled system can do is to request for the
service, the model can avoid making the request by executing the idling
transition.

Because the solution adopted in Section~\ref{S:example} is different, it is
justified to ask whether it is appropriate.
Certain process-algebraic semantic theories provide strong evidence that it is
appropriate.
For the benefit of non-process-algebra-oriented readers, the discussion below
is at the intuitive level.


The standard semantics of CSP~\cite{Ros10}, should testing~\cite{ReV07}, and
the CFFD and NDFD semantics~\cite{VaT95}, among others, naturally yield a
notion for deeming a process ``better than'' or ``as good as'' another
process.
The notion also applies to systems built as compositions of processes.
If a component of a system is replaced by a ``better'' component, then the
system as a whole either becomes ``better'' or remains ``equally good''.
Within the limits of the information that is preserved by the semantics, if a
system satisfies a specification, then also all ``better'' systems satisfy it.
For instance, if a system satisfies a next-state-free linear temporal logic
specification, then also each ``NDFD-better'' and each ``CFFD-better'' system
satisfies it.

\begin{figure}\mbox{}\hfill
\begin{pspicture}(88,31)
\ainit{12,22}
\arc(11.1,20.2)(2.9,3.8)\rput[Br](6,10){$\tau$}
\arc(12.9,20.2)(21.1,3.8)\rput[Bl](18,10){$\tau$}
\arc(24,2)(70,2)\rput[B](47,4){\ts{request}}
\arc(71.1,3.8)(62.9,20.2)\rput[Bl](68,10){\ts{enter}}
\arc(60,22)(14,22)\rput[B](37,24){\ts{leave}}
\psst(2,2){2}\psst(22,2){2}\psst(72,2){2}\psst(12,22){2}\psst(62,22){2}
\end{pspicture}\hfill\mbox{}
\caption{A generic customer proper of a mutual exclusion
system.}\label{F:cust-p}
\end{figure}
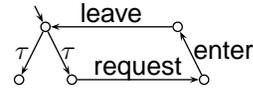

To apply this idea to Peterson-$n$, assume that each customer is split to two
processes, a customer proper and a server.
The customer proper is shown in Figure~\ref{F:cust-p}.
The edges that are labelled with $\tau$ denote activities that are
neither affected by nor observable by the rest of the model.
The server is like in Figure~\ref{F:Pete-q}.
\ts{Request} synchronizes with moving from local state 0 to local state 1
(that is, starting to try to go to the critical section), \ts{enter} with the
arrival to local state 7 (that is, arriving to the critical section), and
\ts{leave} with leaving local state 7.

This customer proper obviously does not do anything that it should not and
does not stop when it should not.
Furthermore, any tentative customer proper that is not ``better than'' or ``as
good as'' Figure~\ref{F:cust-p}, tries to execute \ts{request}, \ts{enter} or
\ts{leave} when it should not; fails to execute \ts{request}, \ts{enter} or
\ts{leave} when it should; or, in the case of CSP, NDFD and CFFD, may steal
all processor time at some point.
So it is unacceptable.
In other words, Figure~\ref{F:cust-p} presents the ``worst'' acceptable
customer proper, the one that makes the greatest challenge to the ability of
the servers to guarantee mutual exclusion and eventual access.
The system is correct with an arbitrary acceptable customer proper if and only
if the system is correct with Figure~\ref{F:cust-p}.
So the customer proper in Figure~\ref{F:cust-p} is most appropriate for a
verification model.

The parallel composition of customer proper with the server yields a process
that plays the role of the original customer.
With Figures~\ref{F:cust-p} and~\ref{F:Pete-q}, the result is like
Figure~\ref{F:Pete-q} with its initial local state replaced by three local
states and two \ttt{goto} commands.
One of the three is the new initial local state, another is a terminal local
state, and the third can be thought of as the initial local state of
Figure~\ref{F:Pete-q}.
The \ttt{goto} commands lead from the new initial local state to the other two
local states.
They do not access any other variables than \ttt{S[i]}.
The atomic operation that models leaving the critical section leads to the new
initial local state instead of the initial local state of
Figure~\ref{F:Pete-q}.

Because the first atomic operation of Figure~\ref{F:Pete-q} does not access
any shared variables, the above-mentioned semantic models see no difference if
it is fused with the \ttt{goto} command from the new initial local state to
the initial local state of Figure~\ref{F:Pete-q}.
Doing so yields precisely the second model of Section~\ref{S:example}.

Further discussion on this issue can be found in~\cite{VaS96}.

\section{Background on Stubborn Sets}\label{S:background}

We will need formal notation for concepts that have been used informally
above.

The set of states is denoted with $S$ and the set of transitions with $T$.
In the case of ASSET, $S$ consists of all possible combinations of values of
the state variables, and $T = \{0, 1, \ldots, |T|-1\}$, where $|T|$ is the
number returned by \ttt{nr\_transitions}.

That $t \in T$ is enabled at $s \in S$ is denoted by $s \ar{t}$, and if
$\neg(s \ar{t})$ holds, then $t$ is disabled at $s$.
The notation $s \arr{t} s'$ denotes that $t$ is enabled at $s$ and if $t$
occurs (that is, is fired) at $s$, then the resulting state is $s'$.
We assume that transitions are \emph{deterministic}.
That is, for any $s \in S$, $s_1 \in S$, $s_2 \in S$, and $t \in T$, if $s
\arr{t} s_1$ and $s \arr{t} s_2$, then $s_1 = s_2$.
A state is \emph{terminal} if and only if no transition is enabled at it.

The obvious extension of $s \arr{t} s'$ to a finite sequence of transitions is
denoted with $s \arr{t_1 \cdots t_n} s'$.
State $s'$ is \emph{reachable} from state $s$ if and only if there is a
(possibly empty) sequence of transitions $t_1 \cdots t_n$ such that $s
\arr{t_1 \cdots t_n} s'$.
The initial state of the model is denoted with $\hat s$.
A state is \emph{reachable} if and only if it is reachable from $\hat s$.

The reachable states and the triples $s \arr{t} s'$ connecting them constitute
a directed graph called \emph{state space}.
For this reason, $s \arr{t} s'$ is called \emph{edge}.
Also other directed graph terminology will be used, such as ``path''.
The state space can be constructed by declaring $\hat s$ as found, and then
firing, in each found state, all the transitions that are enabled in it.
Each resulting state is declared as found.
The algorithm is continued until all found states have been processed.

The \emph{basic strong stubborn set method} assigns to each $s \in S$ a subset
of transitions $\mc{T}(s) \subseteq T$, called \emph{stubborn set}, such that
the following conditions hold.
In them, it is assumed that $t \in \mc{T}(s_0)$ and $t_1 \notin \mc{T}(s_0)$,
\ldots, $t_n \notin \mc{T}(s_0)$.

\begin{itemize}

\item[D0:] If $s_0$ is not terminal, then $\mc{T}(s_0)$ contains an enabled
transition.

\item[D1:] If $s_0 \arr{t_1 \cdots t_n t} s'_n$, then $s_0 \arr{t t_1 \cdots
t_n} s'_n$.

\item[D2:] If $s_0 \ar{t}$ and $s_0 \arr{t_1 \cdots t_n} s_n$, then $s_n
\ar{t}$.

\end{itemize}

Many practical algorithms for constructing sets with the above properties have
been presented.
ASSET uses the strong component algorithm in~\cite{Val88}.
To compute strong components, it uses the optimized version of Tarjan's
algorithm~\cite{Tar72} that has been presented in~\cite{EvK77}.

The \emph{reduced state space} is constructed otherwise like the state space,
but only the enabled transitions in $\mc{T}(s)$ are fired at $s$.
The phrase \emph{full state space} is a synonym of state space.
It is useful when it may be unclear whether ``state space'' refers to the
reduced or full state space.
The sets of states and edges in the reduced state space are obviously subsets
of the sets of states and edges in the full state space.
We say that a state is an \emph{r-state}, an edge is an \emph{r-edge}, a path
is an \emph{r-path}, and so on, if and only if it is in the reduced state
space.
Obviously every r-state is a reachable state and every r-path is a path, but
not necessarily vice versa.

The benefit of stubborn sets is that the reduced state space is often much
smaller than the full state space, so its construction requires less time and
memory.
Even so, it can be used to check many properties of the model.

Letting $s = \hat s$ in the next theorem yields that the reduction preserves
all reachable terminal states of the model, and for each path to a terminal
state, the reduction preserves some permutation of it.
Furthermore, an r-state is terminal if and only if it looks like terminal in
the reduced state space.
Although the theorem is old, its proof is presented here, because it is
essential background to the new results in the next section.

\begin{theorem}[old]\label{T:DL}
Let $s$ be an r-state and $s_\ms{t}$ be a terminal state such that $s \arr{t_1
\cdots t_n} s_\ms{t}$.
The following hold.
\begin{itemize}

\item
There is an r-path from $s$ to $s_\ms{t}$ such that its sequence of
transitions is a permutation of $t_1 \cdots t_n$.

\item
No r-edge starts at $s_\ms{t}$.

\item
If $s$ is an r-state and no r-edge starts at $s$, then $s$ is terminal.

\end{itemize}
\end{theorem}
\begin{IEEEproof}
If $n = 0$, then the first claim is obvious.
Otherwise $s \ar{t_1}$.
By D0, some $t \in \mc{T}(s)$ is enabled at $s$.
If none of $t_1$, \ldots, $t_n$ is in $\mc{T}(s)$, then D2 yields $s_\ms{t}
\ar{t}$, contradicting the assumption that $s_\ms{t}$ is terminal.
So there is $1 \leq i \leq n$ such that $t_i \in \mc{T}(s)$ and none of $t_1$,
\ldots, $t_{i-1}$ is in $\mc{T}(s)$.
By D1, there is $s'$ such that $s \arr{t_i} s' \arr{t_1 \cdots t_{i-1} t_{i+1}
\cdots{t_n}} s_\ms{t}$.
The state $s'$ is an r-state, and there is a path from $s'$ to $s_\ms{t}$ of
length $n-1$.
Repetition of this argument $n$ times yields the first claim.

The second claim follows trivially from the fact that $\mc{T}(s_\ms{t})
\subseteq T$ and $s_\ms{t}$ is terminal.
The last claim follows trivially from D0.
\end{IEEEproof}

Also a lemma will be used in the sequel.
\begin{lemma}\label{L:mid}
If $t \in \mc{T}(s_0)$, $t_1 \notin \mc{T}(s_0)$, \ldots, $t_n \notin
\mc{T}(s_0)$, $s_0 \arr{t_1} s_1 \arr{t_2} \ldots \arr{t_n} s_n$, and $s_0
\arr{t} s'_0 \arr{t_1} s'_1$ $\arr{t_2}\ldots \arr{t_n} s'_n$, then for $1
\leq i \leq n$ we have $s_i \arr{t} s'_i$.
\end{lemma}
\begin{IEEEproof}
Let $1 \leq i \leq n$.
By D2, there is $s''_i$ such that $s_i \arr{t} s''_i$.
By D1, $s_0 \arr{t t_1 \cdots t_i} s''_i$.
Because transitions are deterministic, $s''_i = s'_i$.
\end{IEEEproof}

\section{Stubborn Sets on \tb{AG} \tb{EF} Terminating
Models}\label{S:new-stubborn}

This section is devoted to new results that concern stubborn sets and \tb{AG}
\tb{EF} terminating models.

\begin{theorem}\label{T:AG-EF}
The basic strong stubborn set method preserves the property ``\tb{AG} \tb{EF}
terminating''.
\end{theorem}
\begin{IEEEproof}
Assume first that the property holds on the full state space.
That is, from every reachable state, a terminal state is reachable.
Consider any r-state $s$.
From it there is a path to a terminal state $s_\ms{t}$.
By Theorem~\ref{T:DL}, there is also an r-path from $s$ to $s_\ms{t}$, and
$s_\ms{t}$ is terminal also in the reduced state space.
So the property holds on the reduced state space as well.

Assume now that the full state space does not contain terminal states.
Then by D0, the reduced state space does not contain terminal states either.
So the property holds on neither state space.

Finally, assume that neither of the preceding cases holds.
That is, the full state space contains a state from which no terminal state is
reachable, but $\hat s$ is not such a state.
We have to prove the existence of an r-state $\tilde s$ from which no
r-terminal state is r-reachable.
By Theorem~\ref{T:DL} it suffices to prove that no terminal state is reachable
from $\tilde s$.

For some natural number $n$ and for $0 \leq i \leq n$, we will show the
existence of a transition $t_i$, r-state $s_i$, states $s'_i$, $s''_i$,
$s^\ms{t}_i$, and finite sequences of transitions $\sigma_i$ and $\rho_i$ such
that $s^\ms{t}_i$ is terminal, $s'_i \arr{t_i} s''_i$, $s_i \arr{\sigma_i}
s'_i \arr{\rho_i} s^\ms{t}_i$, and there is no path from $s''_i$ to a terminal
state.
Furthermore, if $i < n$, either $\sigma_{i+1}$ is shorter than $\sigma_i$, or
they are of the same length but $\rho_{i+1}$ is shorter than $\rho_i$.

We choose $s_0 = \hat s$.
Because the first case above does not hold, a state $s'''$ is reachable from
which no terminal state is reachable.
Because the second case above does not hold, a terminal state is reachable
from $s_0$.
Therefore, along the path from $s_0$ to $s'''$ there is an edge $s'_0
\arr{t_0} s''_0$ such that no terminal state is reachable from $s''_0$ and a
terminal state which we call $s^\ms{t}_0$ is reachable from $s'_0$.
Then $\sigma_0$ and $\rho_0$ may be chosen such that $s_0 \arr{\sigma_0} s'_0
\arr{\rho_0} s^\ms{t}_0$.
The base case has been proven.

To prove the induction step, we first consider the case where $t_i \notin
\mc{T}(s_i)$.
Similarly to the proof of Theorem~\ref{T:DL}, an application of D0 and D1 to
the path $s_i \arr{\sigma_i} s'_i \arr{\rho_i} s^\ms{t}_i$ yields $s_{i+1}$,
$s'_{i+1}$, $t'_{i+1} \in \mc{T}(s_i)$, $\sigma_{i+1}$, and $\rho_{i+1}$ such
that $s_i \arr{t'_{i+1}} s_{i+1} \arr{\sigma_{i+1}} s'_{i+1} \arr{\rho_{i+1}}
s^\ms{t}_i$.
Either $\rho_{i+1} = \rho_i$ and $\sigma_{i+1}$ is obtained from $\sigma_i$ by
removing $t'_{i+1}$, or the same holds with the roles of $\sigma$ and $\rho$
swapped.
In the former case, $s'_{i+1} = s'_i$ and we let $s''_{i+1} = s''_i$.
Otherwise by Lemma~\ref{L:mid} $s'_i \arr{t'_{i+1}} s'_{i+1}$, by D2 there is
$s''_{i+1}$ such that $s''_i \arr{t'_{i+1}} s''_{i+1}$, and by D1 and because
$t'_{i+1}$ and $t_i$ are deterministic $s'_{i+1} \arr{t_i} s''_{i+1}$.
No terminal state is reachable from $s''_{i+1}$, because otherwise a terminal
state would be reachable from $s''_i$.
The induction step is completed by choosing $s^\ms{t}_{i+1} = s^\ms{t}_i$ and
$t_{i+1} = t_i$.

The case $t_i \in \mc{T}(s_i)$ remains.
Then D1 can be applied to the path $s_i \arr{\sigma_i} s'_i \arr{t_i} s''_i$.
If it picks a transition from $\sigma_i$, then the case is similar to the case
$\rho_{i+1} = \rho_i$ above.
Otherwise there is an r-state $s_{i+1}$ such that $s_i \arr{t_i} s_{i+1}
\arr{\sigma_i} s''_i$.
If no terminal state is reachable from $s_{i+1}$, then it qualifies as $\tilde
s$.
Otherwise the same reasoning as in the base case with $s_{i+1}$ playing the
role of $\hat s$ and $s''_i$ playing the role of $s'''$ yields $t_{i+1}$,
$s'_{i+1}$, $s''_{i+1}$, $s^\ms{t}_{i+1}$, $\sigma_{i+1}$, and $\rho_{i+1}$
with the required properties.
The length claim holds, because $\sigma_{i+1}$ is a proper prefix of
$\sigma_i$.

Each step of the construction in the proof either yields $\tilde s$, shortens
$\sigma_i$, or shortens $\rho_i$ while retaining the length of $\sigma_i$.
Because $\sigma_i$ and $\rho_i$ cannot become shorter without limit as $i$
grows, eventually $\tilde s$ is obtained.
\end{IEEEproof}

If the reduced state space is constructed in depth-first order, then it is
possible to check efficiently on-the-fly that it is \tb{AG} \tb{EF}
terminating.
By Theorem~\ref{T:AG-EF}, the result applies also to the full state space.
The check is based on computing the strong components of the reduced state
space on-the-fly using Tarjan's algorithm~\cite{Tar72,EvK77}, recognizing
terminal states, and propagating backwards the information whether a terminal
state is reachable.

ASSET works in breadth-first order.
So it cannot use this algorithm.
Instead, it performs the check as a post-processing step.
It re-constructs the edges storing them in reversed direction, and then
performs a linear-time graph search starting at each terminal state.

The idea behind the implementation of stubborn sets in ASSET is that the
modeller should try to make the model \tb{AG} \tb{EF} terminating, but it is
the responsibility of ASSET and not of the modeller to detect if it is not.
The next three theorems list three properties that the basic strong stubborn
set method preserves, if the model indeed is \tb{AG} \tb{EF} terminating.
For all of them, a counterexample found by the method is valid even if the
model is not \tb{AG} \tb{EF} terminating, but if the method finds no
counterexamples, the result can be trusted only if the model is \tb{AG}
\tb{EF} terminating.
Therefore, ASSET first checks the first two of them (the third one has not yet
been implemented).
If it finds no errors, it checks that the model is \tb{AG} \tb{EF}
terminating, giving an error message if it is not.

The following theorem tells that a well-known simple tool for checking
linear-time safety properties works in our context.

\begin{theorem}\label{T:chk_state}
Assume that the model is \tb{AG} \tb{EF} terminating.
For any transition $t_\ms{s}$, the basic strong stubborn set method preserves
the property ``$t_\ms{s}$ may become enabled''.
\end{theorem}
\begin{IEEEproof}
If $t_\ms{s}$ cannot become enabled in the full state space, then clearly it
cannot become enabled in the reduced state space either.
If $t_\ms{s}$ may become enabled in the full state space, then there is a path
$\hat s \arr{t_1 \cdots t_n} s_\ms{t}$ from the initial state to a terminal
state such that $t_\ms{s} = t_i$ for some $1 \leq i \leq n$.
By Theorem~\ref{T:DL}, $t_\ms{s}$ occurs also in the reduced state space.
\end{IEEEproof}

If the construction of the reduced state space is aborted when $t_\ms{s}$ is
found enabled, then $t_\ms{s}$ is never fired.
In that case, $t_\ms{s}$ need not contain statements that change the state;
the enabling condition suffices.
Even so, to use Theorem~\ref{T:chk_state}, $t_\ms{s}$ must be taken into
account in the construction of the stubborn sets.
In ASSET, the enabling condition of $t_\ms{s}$ is represented via the
\ttt{check\_state} function.

To detect complicated errors, additional state variables that store some
information about the history of the execution may be added to the model.
For instance, consider a protocol whose purpose is to deliver messages from a
sending site to a receiving site over an unreliable channel.
To verify the ability of the protocol to prevent distortion of messages, when
a message is given to the protocol, a copy of it is stored in an extra state
variable.
When the protocol delivers the message at the receiving site,
\ttt{check\_state} checks whether it is identical to what was stored in the
extra variable.

The next theorem assumes that each state is classified as a progress state or
other state.
Typically progress states are those where the user either has not requested
for service or has received the service that it requested.
With this convention, a non-progress error occurs if and only if the user has
requested for service but does not get it.

ASSET distinguishes between two types of progress states: \emph{may} and
\emph{must}.
In must progress, \emph{every} path must lead to a progress state.
If the state is terminal, then it must be a progress state in itself.
This is the notion of progress typically used in linear temporal logic.
In may progress, it suffices that \emph{at least one} path leads to a progress
state.
Must progress implies may progress, but not necessarily vice versa.
May progress is a branching-time property and related to the notion of home
properties in Petri nets.
It can be expressed in CTL as \tb{AG} \tb{EF} progress.

For reasons briefly mentioned in Section~\ref{S:intro}, the stubborn set
implementation of ASSET does not support must progress.
The support of may progress is based on the following theorem.

\begin{theorem}
Assume that the model is \tb{AG} \tb{EF} terminating.
Its full state space contains a state from which no progress state is
reachable if and only if it contains a terminal state that is not a progress
state if and only if the reduced state space obtained with the basic strong
stubborn set method contains such a state.
\end{theorem}
\begin{IEEEproof}
Assume that $s$ is reachable but no progress state is reachable from it.
Because the model is \tb{AG} \tb{EF} terminating, a terminal state is
reachable from $s$.
It is a terminal state that is not a progress state.
If a terminal state is not a progress state, then obviously no progress state
is reachable from it.
The first claim has been proven.
The second claim follows from Theorem~\ref{T:DL}.
\end{IEEEproof}

By the theorem, no other support for may progress would be needed in the case
of \tb{AG} \tb{EF} terminating systems than the \ttt{check\_deadlock} feature
of ASSET.
Furthermore, it can be used for all customers simultaneously.
However, when ASSET is used for other kinds of systems without stubborn sets,
the notion of may progress states is useful.
It is convenient that they can also be used with stubborn sets when they work
with them.

The last theorem in this section can be used to check some linear-time
liveness properties, such as ``if the channel of a protocol passes (that is,
does not lose) infinitely many messages, then the protocol as a whole passes
infinitely many messages''.
Actually, it locates the challenge that linear-time liveness causes to
stubborn sets precisely as the problem of preserving cycles that do not make
progress.

\begin{theorem}
Let $t_\omega \in T$ and $T_* \subseteq T$.
The basic strong stubborn set method on \tb{AG} \tb{EF} terminating models
preserves the property ``there is an execution where $t_\omega$ occurs
infinitely many times but no member of $T_*$ occurs infinitely many times''.
\end{theorem}
\begin{IEEEproof}
If such an execution exists in the reduced state space, then it is present
also in the full state space.

Now assume that such an execution exists in the full state space.
It is of the form $\hat s \arr{\rho} s_0 \ar{t_\omega \sigma_1} s_1
\ar{t_\omega \sigma_2} \ldots$, where no element of $T_*$ occurs in any of the
$\sigma_i$ (but $t_\omega$ may occur in $\rho$).
Let $n_\ms{f}$ be the number of states in the full state space, and let $m = 2
n_\ms{f}^2$.
There are $0 \leq j < k \leq n_\ms{f}$ such that $s_j = s_k$.
There are $s_\ms{t}$, $\rho_1$, and $\rho_2$ such that $s_\ms{t}$ is terminal,
$\hat s \arr{\rho_1} s_j \arr{\rho_2} s_\ms{t}$, $|\rho_1| < n_\ms{f}$, and
$|\rho_2| < n_\ms{f}$.

We have $\hat s \arr{\rho_1} s_j \arr{(t_\omega \sigma_{j+1} \cdots t_\omega
\sigma_k)^m} s_j \arr{\rho_2} s_\ms{t}$.
The application of Theorem~\ref{T:DL} to it yields an execution in the reduced
state space that contains at most $2n_\ms{f} - 2$ occurrences of elements of
$T_*$ and at least $2n_\ms{f}^2$ occurrences of $t_\omega$.
It has at least one part that contains at least $n_\ms{f}$ occurrences of
$t_\omega$ and no occurrences of elements of $T_*$.
Because the reduced state space contains no more states than the full state
space, this part contains a cycle.
The prefix of the execution up to the cycle together with an infinite number
of repetitions of the cycle constitutes the infinite execution whose existence
had to be proven.
\end{IEEEproof}

This theorem does not facilitate meaningful use of B\"uchi automata with
stubborn sets, because a B\"uchi automaton observes every action by the
system.
Thus it forces every enabled transition to every stubborn set, so the state
space will not become smaller.
For this reason, a related type of automata has been defined that only
observes actions that may affect the validity of the property~\cite{HPV02}.
Unfortunately, some properties require the detection of cycles consisting
solely of transitions that the automaton does not synchronize with.
This seems to require the linear-time liveness cycle condition and perhaps
also the representation of fairness assumptions as not part of the formula.

\section{Stubborn Sets in the Experiments}\label{S:stubbrule}

The construction of stubborn sets relies on rules of the form ``if this
transition is in the stubborn set of the current state, then also these other
transitions must be''.
A complete implementation of stubborn sets would contain a preprocessor tool
that reasons these rules from the model.
Unfortunately, ASSET is not complete in this respect.
As a consequence, the rules must be provided by the modeller.
This is unacceptable from the point of view of industrial use, but is
sufficient for making scientific experiments.

\begin{figure}{\small
\begin{verbatim}
void next_stubborn( unsigned i ){
  if( i >= n ){ stb(i-n); return; }
  switch( S[i] ){
  case 0: stb(i+n); return;
  case 1: return;
  case 2: stb_all(); return;
  case 3: stb_all(); return;
  case 4:
    if( T[ j[i] ] != i ){ return; }
    stb_all(); return;
  case 5: return;
  case 6: stb_all(); return;
  case 7: stb_all(); return;
  case 8: return;
  default: stb_all(); return;
  }
}
\end{verbatim}}
\caption{Stubborn set rules for three models of
Peterson-$n$.}\label{F:stubbrule}
\end{figure}

Figure~\ref{F:stubbrule} shows the rules used in other experiments of this
publication than the first.
(In the first experiment, an obvious adaptation of the rules was used.)
To discuss them, let $s_0$ denote the current state and $i \leadsto j$ denote
that if transition number $i$ is in the stubborn set $\mc{T}(s_0)$, then ASSET
makes sure that also $j \in \mc{T}(s_0)$.
It was mentioned after Theorem~\ref{T:chk_state} that also the enabling
condition of \ttt{check\_state} must be taken into account.
It will be discussed as a separate case and can be ignored until then.

The discussion below emphasizes the reasons why the rules are valid.
So it gives an over-pessimistic impression of how difficult it is for a human
or preprocessor tool to find the rules.
Excluding case 4, all the rules arise from simple principles.
Even for case 4, it is not beyond imagination that a preprocessor could find
its rule.

Consider first the case $n \leq i < 2n$.
Transition $i$ models customer $i-n$ moving from the initial to the terminal
local state.
Assume that it is the $t$ of D1 and D2.
So it is in $\mc{T}(s_0)$.
The call \ttt{stb(i-n)} makes $i \leadsto i-n$ hold, implying that also
transition $i-n$ is in $\mc{T}(s_0)$.
It models all the remaining atomic operations of customer $i-n$.

If transition $i$ is disabled, then D2 holds trivially.
Furthermore, the only way to enable it is that customer $i-n$ moves to local
state 0.
Therefore, if $s_0 \arr{t_1 \cdots t_n} s_n$ and none of $t_1$, \ldots, $t_n$
is in $\mc{T}(s_0)$, then none of $t_1$, \ldots, $t_n$ is transition $i-n$, so
transition $i$ is disabled at $s_n$.
This implies D1.

Assume now that transition $i$ is enabled.
Clearly the only way to disable it is that either it or transition $i-n$
occurs.
So D2 holds.
Because transition $i$ does not access any variable that the other customers
access, also D1 holds.

The case $0 \leq i < n$ remains.
Transition $i$ is disabled only in case 8.
In that case nothing can enable it, so D1 and D2 hold independently of what
other transitions are in $\mc{T}(s_0)$.
Thus no rule of the form $i \leadsto j$ is needed.
From now on we assume that transition $i$ is enabled.

In cases 0, 1, and 5 transition $i$ does not access variables used by other
customers.
We already say that transition $i+n$ never accesses variables used by other
customers.
So $\mc{T}(s_0) = \{i,i+n\}$ suffices, and no rule of the form $i \leadsto j$
where $j$ refers to another customer is needed to make D1 and D2 hold.
The rule $i \leadsto i+n$ can be dropped in cases 1 and 5, because then
transition $i+n$ is disabled, so it does not matter whether $\mc{T}(s_0) =
\{i,i+n\}$ or $\mc{T}(s_0) = \{i\}$.

In cases 2, 3, and 6 D1 and D2 are forced to hold by introducing a rule of the
form $i \leadsto j$ for every transition $j$.
As a consequence, no transition can be any of the $t_1$, \ldots, $t_n$ of D1
and D2.

The crucial observation behind case 4 is that the only transition that can
turn \ttt{T[j[i]]} \ttt{!=} \ttt{i} from \ttt{true} to \ttt{false} is $i$.
The other customers may write to \ttt{T[j[i]]}, but they write to it their own
index instead of $i$.
Therefore, if \ttt{T[j[i]]} \ttt{!=} \ttt{i}, then D2 holds automatically.
D1 is not a problem either, because other than this test, the atomic operation
does not access variables that are used by other customers.
If \ttt{T[j[i]]} \ttt{==} \ttt{i}, then D1 and D2 are established like in
cases 2, 3, and 6.

Case 7 is affected by \ttt{check\_state}.
(Without it, no rules would be needed.)
Its rule comes from the principle that if a transition may change the state
``further away'' from the checked condition, then rules must be added to a
``sufficient'' subset of transitions that may change the state ``closer to''
the condition.
Full formal treatment of this principle is beyond the scope of this
publication but, to give some idea, let us show the correctness of this
particular rule.

Let $t_\ms{c}$ be an imaginary transition such that its occurrence does not
change the state and it is enabled if and only if the \ttt{check\_state} in
Figure~\ref{F:Pete-q} returns \ttt{true}.
Case 7 may be interpreted as implementing the rules $i \leadsto t_\ms{c}
\leadsto j$ for every transition $j$, forcing D1 and D2 to hold.
It remains to be shown that excluding case 7, $i \leadsto t_\ms{c}$ is not
needed.
That is, we must show that $t_\ms{c}$ may be added to $T \setminus
\mc{T}(s_0)$ without invalidating D1 and D2.

To prove that D2 remains valid, assume that $s_0 \ar{t}$ and $s_0 \arr{t_1
\cdots t_\ms{c} \cdots t_n} s_n$.
Because $t_\ms{c}$ does not change the state, we have $s_0 \arr{t_1 \cdots
t_n} s_n$.
The assumption that D2 was valid beforehand yields $s_n \ar{t}$.

To prove that D1 remains valid, for some $0 \leq i \leq n$ let $s_0 \arr{t_1
\cdots t_i} s_i \arr{t_\ms{c}} s_i \arr{t_{i+1} \cdots t_n} s_n \arr{t} s'_n$.
By D1 there are $s'_0$, \ldots, $s'_{n-1}$ such that $s_0 \arr{t} s'_0
\arr{t_1} s'_1 \arr{t_2}$ $\ldots \arr{t_n} s'_n$.
Lemma~\ref{L:mid} yields $s_i \arr{t} s'_i$.
Because case 7 has been excluded from the discussion, the occurrence of $t$
does not change the value of any \ttt{S[}$i'$\ttt{]} from 7 to something else.
So $s_i \arr{t_\ms{c}}$ implies $s'_i \arr{t_\ms{c}}$.
Because $t_\ms{c}$ does not change the state, we have D1.

\section{Experiments and Discussion}\label{S:conclusions}

\begin{table}
\begin{tabular}{@{}r|rrr|rrr@{}}
    & \multicolumn{3}{c|}{plain} & \multicolumn{3}{c}{stubborn sets} \\
$n$ & states & edges & time & states & edges & time\\
\hline
  & \multicolumn{6}{c}{~ ~ not non-progress revealing}\\
2 &          133 &           266 &  0.0 &          88 &          124 &  0.1 \\
3 &      38\,038 &      114\,114 &  0.3 &     18\,817 &      34\,083 &  0.2 \\
4 & 12\,346\,971 &  49\,387\,884 & 70.3 & 4\,312\,993 &  8\,988\,034 & 22.2 \\
\hline
  & \multicolumn{6}{c}{~ ~ non-progress revealing}\\
2 &          163 &           326 &  0.1 &         116 &          162 &  0.0 \\
3 &      43\,675 &      131\,025 &  0.3 &     23\,134 &      41\,562 &  0.2 \\
4 & 14\,186\,506 &  56\,746\,024 & 85.6 & 5\,316\,461 & 10\,903\,336 & 36.9 \\
\hline
  & \multicolumn{6}{c}{~ ~ correct}\\
2 &          574 &        1\,148 &  0.0 &         378 &          522 &  0.0 \\
3 &      96\,854 &      290\,562 &  0.4 &     44\,868 &      78\,750 &  0.3 \\
4 & 26\,209\,918 & 104\,839\,672 &  184 & 9\,318\,636 & 18\,581\,236 & 62.7 \\
\hline
  & \multicolumn{6}{c}{~ ~ mutex-violating}\\
2 &          336 &           602 &  0.0 &         219 &          258 &  0.0 \\
3 &      32\,957 &       87\,081 &  0.2 &     15\,164 &      22\,100 &  0.2 \\
4 &  6\,614\,675 &  23\,547\,787 & 16.4 & 2\,116\,738 &  3\,527\,255 &  5.8 \\
\end{tabular}
\caption{Results with ASSET on models of Peterson-$n$.}\label{T:result}
\end{table}

Table~\ref{T:result} shows, for the models discussed in this publication and
for different values of $n$, the number of reachable states, the number of
edges in the state space (that is, successful transition firings), and the
time it took to construct and explore the state space.
The time is in seconds.
In addition to the time in the table, a couple of seconds were spent on each
model by the C++ compiler.
The experiment was made on a Linux 1.6~GHz dual-core laptop with 2~GB of
memory.

Because a safety error was detected in the mutex-violating model, the
postprocessing steps that check progress properties and \tb{AG} \tb{EF}
termination were not executed.
This explains the exceptionally short times obtained with the model.

A comparison of the results on the first two models tells that the addition of
terminal states and transitions to them did not make the state space grow
much.

ASSET has also an implementation of the well-known symmetry reduction method.
However, the models discussed in this publication are not symmetric.
Experiments have also been made with models where the $\forall$ test is
represented as a single atomic operation.
The symmetry method can then be used.
Both methods together reduced the number of states of a deadlocking version to
quadratic in $n$.


\section*{Acknowledgement}

I thank the anonymous reviewers for helpful comments.


\end{document}